\pdfoutput=1

\documentclass[11pt]{article}

\usepackage[final]{acl}

\usepackage{times} 
\usepackage{latexsym}
\usepackage{xspace}

\newcommand*\circled[1]{\tikz[baseline=(char.base)]{
            \node[shape=circle,draw,inner sep=.6pt] (char) {#1};}}

\usepackage[T1]{fontenc}

\usepackage[utf8]{inputenc}

\usepackage{microtype}

\usepackage{inconsolata}

\usepackage{pgfplots}

\usepackage{paralist}
\usepackage{xcolor}

\usepackage[most]{tcolorbox}
\usepackage{colortbl}
\usepackage{enumitem}
\newtcolorbox{mybox}[2][]{
    colback=white,
    colframe=green!45,
    fonttitle=\bfseries,
    coltitle=black,
    sharp corners,
    title=#2,
    #1
}

\definecolor{squadbase}{HTML}{0073C2}
\newcommand{\squadcolor}[1]{\setlength{\fboxsep}{1.5pt}\colorbox{squadbase!25}{#1}}

\definecolor{hotpotbase}{HTML}{EFC000}
\newcommand{\hotpotcolor}[1]{\setlength{\fboxsep}{1.5pt}\colorbox{hotpotbase!35}{#1}}

\definecolor{quacbase}{HTML}{00A087}
\newcommand{\quaccolor}[1]{\setlength{\fboxsep}{1.5pt}\colorbox{quacbase!25}{#1}}

\definecolor{coqabase}{HTML}{CD534C}
\newcommand{\coqacolor}[1]{\setlength{\fboxsep}{1.5pt}\colorbox{coqabase!25}{#1}}

\definecolor{doqabase}{HTML}{FF8C38}
\newcommand{\doqacolor}[1]{\setlength{\fboxsep}{1.5pt}\colorbox{doqabase!25}{#1}}

\definecolor{orquacbase}{HTML}{2ECC71}
\newcommand{\orquaccolor}[1]{\setlength{\fboxsep}{1.5pt}\colorbox{orquacbase!25}{#1}}

\definecolor{syntheticbase}{HTML}{9B4DCA}
\newcommand{\syntheticcolor}[1]{\setlength{\fboxsep}{1.5pt}\colorbox{syntheticbase!25}{#1}}

\usepackage{booktabs}
\usepackage{multirow}

\newcommand{\synqa}{\textsc{SynQA}\xspace}
\newcommand{\synatt}{\textsc{Syn-Att}\xspace}

\usepackage{graphicx}
\usepackage{subcaption}

\usepackage{booktabs} 

\usepackage{multirow}

\usepackage{tcolorbox}

\newtcolorbox{prompt}{
    colback=black!3,
    colframe=black!40,
    boxrule=0.5pt,
    left=8pt,
    right=8pt,
    top=8pt,
    bottom=8pt,
    arc=2pt,
    breakable,
    enhanced,
    before skip=10pt,
    after skip=10pt
}

\title{On Synthesizing Data for Context Attribution in Question Answering}

\author{
\textbf{Gorjan Radevski}$^{5,6}$\thanks{Both authors contributed equally to this work.\\For correspondence reach out to: gorjan.radevski@gmail.com, kiril.gashteovski@neclab.eu, carolin.lawrence@neclab.eu}, 
\textbf{Kiril Gashteovski}$^{1,8}$\footnotemark[1], 
\textbf{Shahbaz Syed}$^1$, 
\textbf{Christopher Malon}$^2$, \\
\textbf{Sebastien Nicolas}$^1$, 
\textbf{Chia-Chien Hung}$^1$, 
\textbf{Timo Sztyler}$^1$, 
\textbf{Verena Heußer}$^1$, \\
\textbf{Wiem Ben Rim}$^4$, 
\textbf{Masafumi Enomoto}$^3$, 
\textbf{Kunihiro Takeoka}$^3$, 
\textbf{Masafumi Oyamada}$^3$, \\
\textbf{Goran Glavaš}$^7$, 
\textbf{Carolin Lawrence}$^1$ \\
$^1$NEC Laboratories Europe, Germany
$^2$NEC Laboratories America, USA \\
$^3$NEC Corporation, Japan
$^4$University College London, UK \\
$^5$KU Leuven, Belgium
$^6$Epimind, Belgium \\
$^7$Center for Artificial Intelligence and Data Science, University of Würzburg, Germany \\
$^8$CAIR, Ss. Cyril and Methodius University of Skopje, North Macedonia \\
}

\begin{document}
\maketitle





\begin{abstract}

Question Answering (QA) accounts for a significant portion of LLM usage ``in the wild''. However, LLMs sometimes produce false or misleading responses, also known as \textit{hallucinations}. Therefore, grounding the generated answers in contextually provided information---i.e., providing evidence for the generated text---is paramount for LLMs' trustworthiness. Providing this information is the task of \textit{context attribution}. In this paper, we systematically study LLM-based approaches for this task, namely we investigate (i) zero-shot inference, (ii) LLM ensembling, and (iii) fine-tuning of small LMs on synthetic data generated by larger LLMs. Our key contribution is \synqa: a novel generative strategy for synthesizing context attribution data. Given selected context sentences, an LLM generates QA pairs that are supported by these sentences. This leverages LLMs’ natural strengths in text generation while ensuring clear attribution paths in the synthetic training data. We show that the attribution data synthesized via \synqa is highly effective for fine-tuning small LMs for context attribution in different QA tasks and domains. Finally, with a user study, we validate the usefulness of small, efficient LMs (fine-tuned on synthetic data from \synqa) in context attribution for QA. 

\end{abstract}

\section{Introduction}





\begin{figure}
    \centering
    \includegraphics[width=1.0\linewidth]{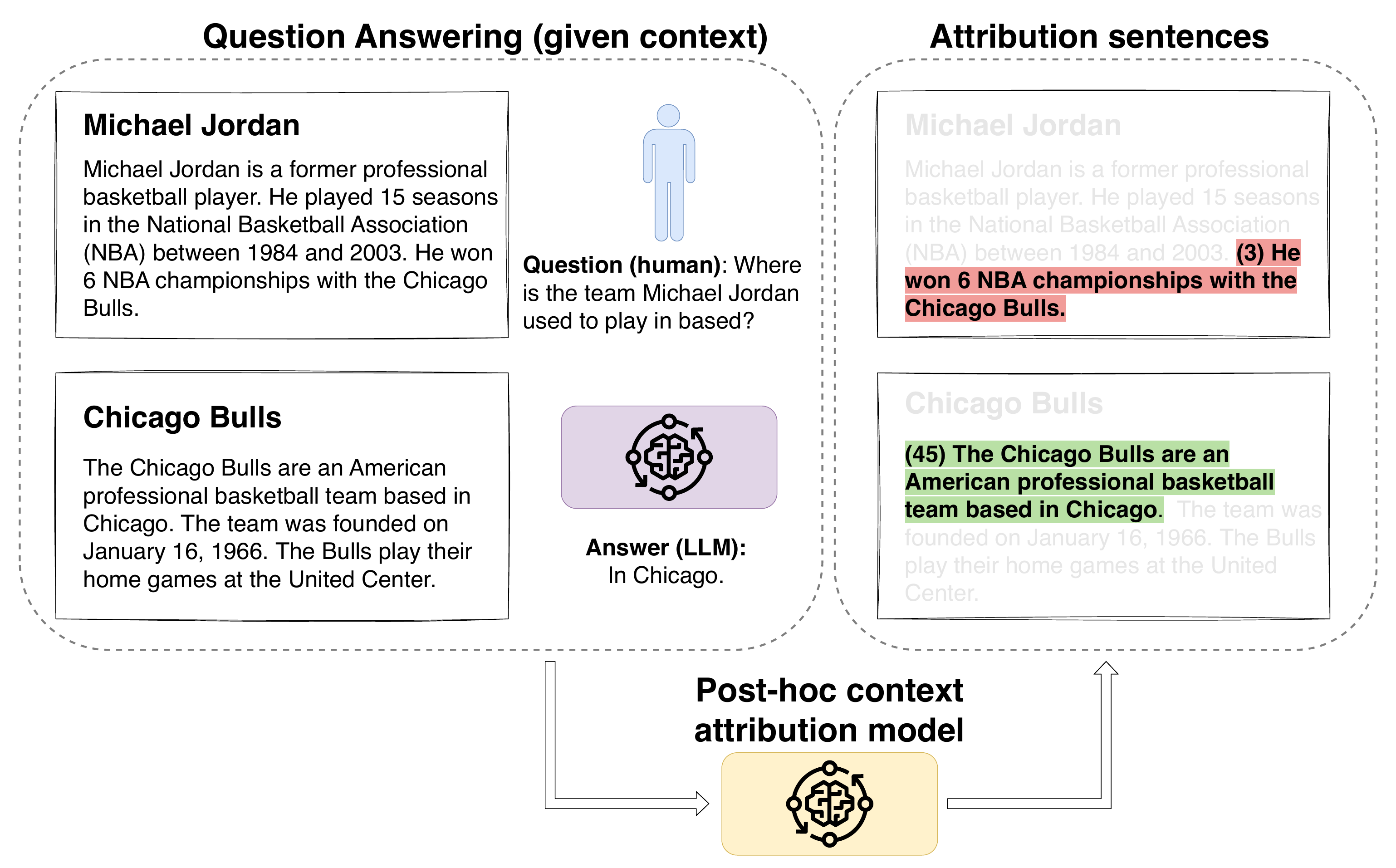}
    \caption{Post-hoc context attribution: Given a question, an LLM-generated answer, and context (from human input or retrieval), the model identifies supporting sentences within the context. Our user study (\S\ref{sec:user-study}) shows that presenting these supporting sentences helps users verify LLM answers more quickly and accurately.}
    \label{fig:fig1}
\end{figure}

Large Language Models (LLMs) have become ubiquitous, with Question Answering (QA) as their most common use case \cite{trippas2024users}. 
However, LLMs have a tendency to hallucinate: they generate content that is factually incorrect w.r.t.~a previously provided reference text. 
This poses the need for \textit{context attribution} methods that create links between the answer and different relevant parts of the (potentially large) reference text; for an illustration of the task, see Figure~\ref{fig:fig1}. 

For these reasons, reliable and efficient context attribution is instrumental in manually verifying the factuality of LLM-generated content. By conducting a user study, \citet{slobodkin2024attribute} report two important findings in this respect: (1) attribution models reduce the human workload in a fact-checking task by as much as 50\%; and (2) sentence level granularity---i.e., grounding the answers in one or more relevant sentences in the reference context---is the most efficient granularity level for manual fact-checking of LLM-generated answers.



%

Given the task importance, recent context attribution research spans text summarization \cite{krishna2023longeval,ernst2024power}, citation attribution \cite{gao2023enabling,Huang2024AdvancingLL}, and question answering \cite{Phukan2024PeeringIT,CohenWang2024ContextCiteAM}. However, the solutions rely on document- or paragraph-level evidence, which comes with the following limitations: (1) the user still has to read the (potentially long) document(s) to verify the generated text; and (2) the LLM needs to correctly generate the reference output alongside answering the question correctly. 
In contrast, the post-hoc attribution methods perform the attribution \emph{after} the LLM generates the answer. These models, however, either attribute to coarse-grained units of text \cite{Nakano2021WebGPTBQ,Menick2022TeachingLM,Buchmann2024AttributeOA}, or provide fine-grained attributions but their inference is computationally expensive \cite{CohenWang2024ContextCiteAM}, hindering their adoption in practice.

In this paper, we explore how LLMs can generate synthetic data for attribution fine-tuning, enabling accurate, sentence-level, and real-time efficient models. For data generation, we compare two approaches: (1) in the fairly straightforward \textit{attribution synthesis} (\synatt), we start with question-answer pairs from a reference text and prompt the LLM to identify the supporting sentences; (2) in our novel \textit{question-answer synthesis} (\synqa), we use Wikipedia sentences and prompt the LLM to generate a question-answer pair that is fully supported by these sentences. Given the generated data, we fine-tune smaller, more efficient context attribution models and compare their performance.

Through extensive evaluation encompassing six datasets and two real-world scenarios (attribution for single-turn questions: i.e., a single question and single answer, and for dialogue questions: i.e., as part of a conversation), we demonstrate that models trained on synthetic data generated by \synqa: \circled{1} Outperform zero-shot LLMs that are orders of magnitude larger, while maintaining real-time inference capabilities (\S\ref{sec:experiments-zero-shot}); \circled{2} Achieve competitive performance on in-domain tasks and superior generalization to out-of-domain datasets compared to models trained on gold data (\S\ref{sec:experiments-gold}); \circled{3} Successfully handle dialogue-based attribution without requiring in-domain training data (\S\ref{sec:experiments-dialog}); \circled{4} Show consistent performance improvements as synthetic training data increases (\S\ref{sec:scalling-trends}); \circled{5} Significantly improve users' speed and accuracy in verifying LLM-generated answers (\S\ref{sec:user-study}). 

These findings suggest that \synqa reduces dependence on large human-labeled datasets while improving context attribution robustness. Our user study further validates the practical utility of fine-tuned small models in real-world question-answering applications, demonstrating their effectiveness in diverse settings. Overall, these results highlight the viability of scalable, data-efficient context attribution techniques, paving the way for more interpretable and trustworthy AI systems.

\section{Synthesizing Attribution Data}

\begin{figure*}[ht]
    \centering
    \includegraphics[width=\textwidth]{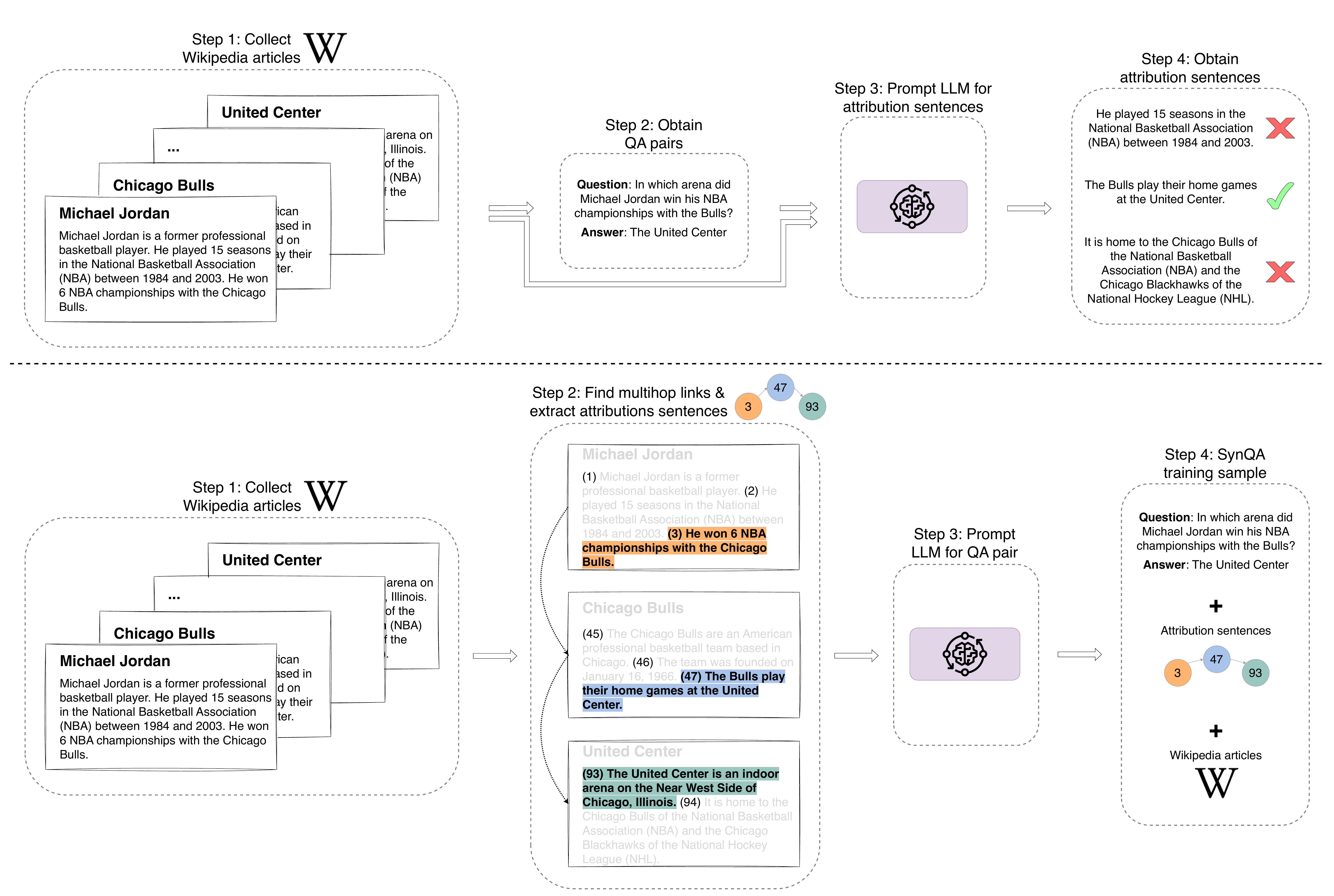}
    \caption{\textbf{Top:} The \synatt baseline method for synthetic attribution data generation. Given context and question-answer pairs, we prompt an LLM to identify supporting sentences, which are then used to train a smaller attribution model. However, this discriminative approach may yield noisy training data as LLMs are less suited for classification tasks (see \S\ref{sec:experiments-zero-shot}). \textbf{Bottom:} The \synqa data generation pipeline leverages LLMs' generative strengths through four steps: (1) collection of Wikipedia articles as source data; (2) extraction of context attributions by creating chains of sentences that form hops between articles; (3) generation of QA pairs by prompting an LLM with only these context attribution sentences; (4) compilation of the final training samples, each containing the generated QA pair, its context attributions, and the original articles enriched with related distractors.}
    \label{fig:method}
\end{figure*}

Context attribution identifies which parts of a reference text support a given question-answer pair~\cite{rashkin2023measuring}. Formally, given a question $q$, its answer $a$, and a context text $c$ consisting of sentences ${s_1, ..., s_n}$, the task is to identify the subset of sentences $S \subseteq c$ that fully support the answer $a$ to question $q$. To train efficient attribution models without requiring expensive human annotations, we explore synthetic data generation approaches using LLMs.
We implement two methods for synthetic data generation. The baseline method (\synatt) is discriminative: given existing question-answer pairs and their context, an LLM identifies supporting sentences (i.e., synthetic context attributions), which are then used to train a smaller attribution model using knowledge distillation. Our proposed method (\synqa) takes a generative approach: given selected context sentences, an LLM generates question-answer pairs that are fully supported by these sentences. This approach better leverages LLMs' natural strengths in text generation while ensuring clear attribution paths in the synthetic training data.





\subsection{\synqa: Generative Synthetic Data Generation Method}\label{sec:synqa_method}

\synqa consists of three parts: context selection, QA generation, and distractors mining (for an illustration of the method, see Figure~\ref{fig:method}). In what follows, we describe each part in detail.

\textbf{Context Collection.} We use Wikipedia as our data source, as each article consists of sentences about a coherent and connected topic, with two collection strategies. In the first, we select individual Wikipedia articles for dialogue-centric generation and use their sentences as context. In the second, for multi-hop reasoning, we identify sentences containing Wikipedia links and follow these links to create ``hops'' between articles, limiting to a maximum of two paths to maintain semantic coherence, while enabling more complex reasoning patterns. See Appendix~\ref{app:data-collection} for more details.

\textbf{Question-Answer Generation.} Given the set of contexts, an LLM (Llama 70B in our implementation) can now generate question-answer pairs. For single articles, we prompt the model to generate
a set of dialogue-centric question-answer pairs, where questions build upon the previous context. For linked articles, we prompt the model to generate questions that necessitate connecting information across the articles, encouraging multi-hop reasoning. Importantly, we provide the LLM with the complete multi-hop reasoning chain as ground truth attribution sentences and ask it to generate question-answer pairs that can only be answered using this evidence.
This yields multi-hop samples requiring integration of information across documents.
We provide more details in Appendix~\ref{app:qa-generation}, and the prompts we use for generating the synthetic data in Appendix \ref{app:prompts}.

\textbf{Distractors Mining.} To make the attribution task more realistic, we augment each sample with distractor articles. With E5 \cite{wang2022text}, we embed each Wikipedia article in our collection. For each article in the training sample, we randomly select up to three distractors with the highest semantic similarity to the source articles. These distractors share thematic elements with the source articles, but lack information to answer the questions. Since Wikipedia articles are unique within a single dump, the source article itself is never included among the distractors. The only scenario where partial information might exist would be if a question addresses a general fact that appears in multiple related articles—a rare occurrence given Wikipedia's article structure and our question generation process. See Appendix~\ref{app:distractors-mining} for more details.


\subsection{Advantages of \synqa}
The \synqa approach has three key advantages:
(1) it leverages LLMs' strength in generation rather than classification; (2) creates diverse training samples requiring both dialogue understanding and multi-hop reasoning; and (3) ensures generated questions have clear attribution paths since they are derived from specific context sentences.
By generating both entity-centric and dialogue-centric samples, \synqa produces training data that reflects the variety of real-world QA scenarios, helping models develop robust attribution capabilities, which our experiments demonstrate to generalize across different contexts and domains.

\section{Experimental Study}\label{sec:experiments}
We conduct a comprehensive evaluation across multiple aspects: comparison with zero-shot LLMs, comparison with models trained on gold attribution data, and generalization to dialogue settings.
With our experiments, we shed light on the performance and practical utility of our approach.




\subsection{Experimental Setting}
We evaluate model performance using precision (P), recall (R), and F1 score. For each sentence in the LLM's output, the context attribution models identify the set of context sentences that support that output sentence. Precision measures the proportion of predicted attributions that are correct, while recall measures the proportion of ground truth attributions that are successfully identified.

For a fair and comprehensive evaluation, we train all models with a single pass over the training data unless stated otherwise, referring to this setup as \textbf{1P} when needed. For a more controlled comparison, some experiments limit the number of training samples each model encounters. Since the synthetic dataset contains approximately 1.0M samples, we allow models to \textit{observe} an equivalent number of samples from the gold training set, ensuring comparable exposure to models trained on data from \synqa. We refer to this setting as \textbf{1M} when necessary. For all models, we fine-tune only the LoRA parameters (alpha=64, rank=32) using a fixed learning rate of 1e-5 and a weight decay of 1e-3. 

\textbf{In-domain datasets:} We use \squadcolor{SQuAD} \cite{Rajpurkar2016SQuAD1Q} and \hotpotcolor{HotpotQA} \cite{Yang2018HotpotQAAD} as our primary in-domain benchmarks.\footnote{For some experiments (e.g., in Table~\ref{table:zero-shot-models}), these datasets are also \textit{out-of-domain} w.r.t. data generated by \synqa.} SQuAD provides clear sentence-level evidence for answering questions, serving as a strong baseline for direct attribution. HotpotQA introduces multi-hop reasoning, requiring models to link information across multiple sentences (sometimes from different articles) to identify the correct evidence chain. Additionally, HotpotQA includes distractor documents—closely related yet incorrect sources—posing a more challenging but realistic setting for evaluating attribution performance.

\textbf{Out-of-domain datasets:} To assess generalization beyond the training distribution, we evaluate models on \quaccolor{QuAC} \cite{Choi2018QuACQA}, \coqacolor{CoQA} \cite{Reddy2018CoQAAC}, \orquaccolor{OR-QuAC} \cite{qu2020open}, and \doqacolor{DoQA} \cite{campos-etal-2020-doqa}. 
These datasets present conversational QA scenarios that differ from SQuAD and HotpotQA. Specifically, QuAC and CoQA introduce multi-turn dialogue structures with coreferences, challenging models to track context across multiple turns. This conversational nature creates a methodological challenge: while these datasets are valuable for evaluating dialogue-based attribution, their reliance on conversation history makes direct comparison with models trained on single-turn QA datasets impossible.

To enable comprehensive evaluation across dialogue QA and single-turn QA, we create two versions from Quac and CoQA:
\begin{inparaenum}[(i)]
    \item a rephrased version using Llama 70B \cite{Dubey2024TheL3} that converts questions into a standalone format for fair comparison with models trained on single-turn context attribution (suffixed by ``-ST''), and
    \item the original (unchanged) version for assessing dialogue-based attribution.
\end{inparaenum}
See Appendix~\ref{app:multi_to_single} for examples.


DoQA extends this challenge further by incorporating domain-specific dialogues (cooking, travel, and movies)
, thus testing the models' adaptability to specialized contexts. OR-QuAC includes 
context-independent rewrites of the dialogue questions, such that they can be posed in isolation of prior context (i.e., single-turn QA). This enables us to test the models on their capabilities in both single-turn QA and dialogue QA settings.

\subsection{Methods}
We compare models trained with data from \synqa against several baselines, including sentence encoder-based models, zero-shot instruction-tuned LLMs, and models trained on synthetic and gold context attribution data. Specifically, we experiment with the following methods:

\paragraph{Sentence-Encoders:} We embed each sentence in the context along with the question-answer pair, and select attribution sentences (from the context) based on the cosine similarity with a fixed threshold, tuned on a small validation set.

\paragraph{Zero-shot LLMs:} We evaluate various instruction-tuned LLMs in a zero-shot manner, as such models have been shown to perform well across a range of NLP tasks \cite{shu2023exploitability,zhang2023instruction}. During inference, we provide an instruction template describing the task to the LLM (see Appendix~\ref{app:zero-shot-prompts} for details).

\paragraph{Ensembles of LLMs:} We aggregate the predictions of multiple LLMs through majority voting, selecting attribution sentences that receive consensus from at least 50\% of the ensemble. In our experiments, we use Llama8B \cite{Dubey2024TheL3}, Mistral7B, and Mistral-Nemo12B \cite{Jiang2023Mistral7} as the ensemble constituents.

\paragraph{Models trained on in-domain gold data:} Fine-tuning on gold-labeled attribution data provides an upper bound on in-domain performance, helping us assess how well synthetic training data generalizes.

\paragraph{\synatt:} \synatt generates synthetic training data by prompting multiple LLMs to perform context attribution in a discriminative manner, aggregating their outputs via majority voting, and training a smaller model on the resulting dataset. To make it a stronger baseline against \synqa, we give the training data of SQuAD and HotpotQA (the context, questions, and answers) to the LLMs and ask them to perform context attribution (note that we do not use the gold attribution). Finally, we train a model on the generated synthetic data---the context, questions, and answers from the training dataset, and the synthetic context attribution links.

\paragraph{\synqa:} We train models using synthetic data generated by our proposed method (\synqa). Importantly,
we ensure that the models are not exposed to \textit{any} part of the evaluation data.\footnote{We identify data leakage by representing each Wikipedia article as a MinHash signature. Then, for each training Wikipedia article, we retrieve candidates from the testing datasets via Locality Sensitivity Hashing and compute their Jaccard similarity \cite{dasgupta2011fast}. We flag as potential leaks pairs exceeding a threshold empirically set to 0.8.}

\subsection{Results and Discussion}
Evaluating our context attribution models requires a multifaceted approach, as performance is influenced by both the quality of training data and the model’s ability to generalize beyond in-domain distributions. Therefore, we design our experiments to address five core questions:
\begin{inparaenum}[(i)]
    \item How well do zero-shot LLMs perform on context attribution QA tasks (\S\ref{sec:experiments-zero-shot})?
    \item Can models trained on synthetic data generated by \synqa exceed the performance of models trained on gold context attribution data (\S\ref{sec:experiments-gold})?
    \item To what extent do models generalize to dialogue settings where in-domain training data is unavailable (\S\ref{sec:experiments-dialog})?
    \item How well do models scale in terms of synthetic data quantity generated by \synqa (\S\ref{sec:scalling-trends})?
    \item How do improved context attributions impact the end users' speed and ability to verify questions answering outputs (\S\ref{sec:user-study})?
\end{inparaenum}


\subsubsection{Comparison to Zero-Shot Models}\label{sec:experiments-zero-shot}

\begin{table*}[t]
\centering
\resizebox{1.0\textwidth}{!}{
\begin{tabular}{lccccccccccccccc} \toprule
\multirow{2}{*}{Model} & \multirow{2}{*}{Training data} & \multicolumn{3}{c}{\squadcolor{SQuAD}} & \multicolumn{3}{c}{\hotpotcolor{HotpotQA}} & \multicolumn{3}{c}{\quaccolor{Quac-ST}} & \multicolumn{3}{c}{\coqacolor{CoQA-ST}} \\ \cmidrule(lr){3-5} \cmidrule(lr){6-8} \cmidrule(lr){9-11} \cmidrule(lr){12-14}
& & P & R & F1 & P & R & F1 & P & R & F1 & P & R & F1 \\ \midrule
\textbf{\textit{Baselines}} \\
Random & -- & 19.8 & 15.4 & 17.3 & 4.8 & 15.2 & 7.3 & 5.2 & 15.1 & 7.7 & 7.3 & 15.1 & 9.9 \\
E5 | 561M & Zero-shot & 38.1 & 76.5 & 50.9 & 12.4 & 41.4 & 19.1 & 65.0 & 73.8 & 69.1 & 61.1 & 15.2 & 24.4 \\
HF-SmolLM2 | 365M & Zero-shot & 28.1 & 46.4 & 35.0 & 5.1 & 7.3 & 6.0 & 10.6 & 22.6 & 14.4 & 10.6 & 21.5 & 14.2 \\
Llama | 1B & Zero-shot & 37.5 & 62.0 & 46.7 & 5.3 & 28.1 & 8.9 & 8.8 & 65.4 & 15.4 & 11.9 & 52.8 & 19.4 \\
Mistral | 7B & Zero-shot & 71.5 & 94.4 & 81.4 & 42.9 & 42.7 & 42.8 & 63.2 & 88.6 & 73.8 & 59.0 & 72.2 & 64.9 \\
Llama | 8B & Zero-shot & 71.9 & 96.9 & 82.6 & 49.2 & 52.9 & 51.0 & 64.1 & 92.1 & 75.6 & 55.7 & 76.4 & 64.4 \\
Mistral-NeMo | 12B & Zero-shot & 89.5 & 94.5 & 91.8 & 46.4 & 47.3 & 46.8 & 81.8 & 85.3 & 83.5 & 79.0 & 67.2 & 72.6 \\
Ensemble | 27B & Zero-shot & 83.1 & 96.3 & 89.2 & 48.1 & 59.6 & 53.2 & 74.8 & 90.3 & 81.8 & 69.5 & 73.6 & 71.5 \\
Llama | 70B & Zero-shot & 95.3 & 95.6 & 95.5 & 87.6 & 37.5 & 52.5 & 89.7 & 87.8 & 88.7 & \textbf{87.5} & \textbf{73.3} & \textbf{79.8} \\
\midrule
\textbf{\textit{Baselines}} \\
Llama | 1B & \synatt (1P) & 89.8 & 96.5 & 93.0 & 50.6 & 58.6 & 54.3 & 64.9 & 91.5 & 75.9 & 53.1 & 75.5 & 62.3 \\
Llama | 1B & \synatt (1M) & 84.3 & \textbf{96.9} & 90.2 & 54.4 & 58.0 & 56.1 & 63.4 & \textbf{92.4} & 75.2 & 52.5 & 77.5 & 62.6 \\ \midrule
\textbf{\textit{Ours}} \\
Llama | 1B & \syntheticcolor{\synqa} & \textbf{96.0} & 96.2 & \textbf{96.1} & \textbf{89.6} & \textbf{69.4} & \textbf{78.2} & \textbf{93.3} & 89.1 & \textbf{91.1} & \underline{82.3} & 68.5 & \underline{74.8} \\
\bottomrule
\end{tabular}
}
\caption{
Performance comparison between zero-shot models and models trained with synthetic data. While larger zero-shot language models perform well, our smaller \synqa model achieves the highest F1 scores across all tasks. \textbf{Bold} indicates the best-performing method; \underline{underlined} indicates our method when it ranks second. 1P = models trained with a single pass over training data; 1M = models trained with 1M samples (matched to \synqa data size).
}
\label{table:zero-shot-models}
\end{table*}

In Table~\ref{table:zero-shot-models}, we present the performance of zero-shot models and those trained without gold context attribution data. State-of-the-art sentence-encoder models (e.g., E5) perform relatively poorly, consistent with prior findings \cite{CohenWang2024ContextCiteAM}. In contrast, LLMs exhibit strong performance, with improvements correlating with model size. Ensembling multiple zero-shot LLMs leverages complementary strengths and further enhances performance, but makes the attribution more expensive.

Models trained using \synqa data outperform all zero-shot baselines except on CoQA. This exception is particularly noteworthy given the nature of CoQA: while both QuAC(-ST) and CoQA(-ST) are conversational QA datasets (different from SQuAD and HotpotQA), CoQA is derived from diverse sources.\footnote{Children's stories (MCTest), classic literature (Project Gutenberg), high school English exams, and news articles.} On HotpotQA, we observe that Llama 70B exhibits high precision (87.6\%), indicating that it rarely selects irrelevant sentences as supporting evidence. However, Llama 1B trained with \synqa data achieves even higher precision (89.6\%), suggesting that \synqa further refines evidence selection. More importantly, \synqa-trained models exhibit higher recall compared to zero-shot LLMs, indicating they retrieve more complete supporting evidence for the QA pairs. This pattern demonstrates that \synqa enables models to identify more comprehensive supporting evidence while maintaining strong precision.

We also tested models trained with the discriminative method \synatt. These models significantly outperform their non-fine-tuned same size counterparts. However, as postulated, our generative approach \synqa outperforms \synatt significantly in all cases as per F1 score. Additionally, \synqa surpasses zero-shot LLMs that are orders of magnitude larger, showing that we can train a model that is both more accurate and efficient.

\subsubsection{Comparison to Models Trained on Gold Attribution Data}\label{sec:experiments-gold}

\begin{table*}[t]
\centering
\resizebox{1.0\textwidth}{!}{
\begin{tabular}{lccccccccccccccc} \toprule
\multirow{3}{*}{Model} & \multirow{3}{*}{Training data} 

& \multicolumn{6}{c}{\textbf{In-Domain}} 
& \multicolumn{6}{c}{\textbf{Out-of-Domain}} \\ \cmidrule(lr){3-8} \cmidrule(lr){9-14}

& & \multicolumn{3}{c}{\squadcolor{SQuAD}} & \multicolumn{3}{c}{\hotpotcolor{HotpotQA}} 
& \multicolumn{3}{c}{\quaccolor{QuAC-ST}} & \multicolumn{3}{c}{\coqacolor{CoQA-ST}} \\ \cmidrule(lr){3-5} \cmidrule(lr){6-8} \cmidrule(lr){9-11} \cmidrule(lr){12-14}

& & P & R & F1 & P & R & F1 & P & R & F1 & P & R & F1 \\ \midrule
\textbf{\textit{Baselines}} \\
Llama | 1B & Zero-shot & 37.5 & 62.0 & 46.7 & 5.3 & 28.1 & 8.9 & 8.8 & 65.4 & 15.4 & 11.9 & 52.8 & 19.4 \\
Llama | 1B & \squadcolor{SQuAD} (1P) & 98.4 & 98.4 & 98.4 & 48.7 & 20.0 & 28.4 & 92.6 & 85.8 & 89.0 & 79.9 & 64.3 & 71.2 \\
Llama | 1B & \hotpotcolor{HotpotQA} (1P) & 41.3 & 87.3 & 56.0 & 87.5 & 79.9 & 83.5 & 45.2 & 89.9 & 60.1 & 41.0 & 70.9 & 52.0 \\
Llama | 1B & \squadcolor{SQuAD} \& \hotpotcolor{HotpotQA} (1P) & 98.3 & 98.3 & 98.3 & \textbf{89.7} & 78.9 & 84.0 & 90.4 & 90.0 & 90.2 & 83.1 & 68.0 & 74.8 \\
Llama | 1B & \squadcolor{SQuAD} \& \hotpotcolor{HotpotQA} (1M) & \textbf{98.3} & \textbf{98.4} & \textbf{98.3} & 87.0 & \textbf{85.2} & \textbf{86.1} & 84.0 & 89.2 & 86.6 & 79.2 & 66.4 & 72.2 \\ \midrule
\textbf{\textit{Ours}} \\
Llama | 1B & \syntheticcolor{\synqa} & 96.0 & 96.2 & 96.1 & \underline{89.6} & 69.4 & 78.2 & \underline{93.3} & 89.1 & \underline{91.1} & 82.3 & 68.5 & \underline{74.8} \\
Llama | 1B & \syntheticcolor{\synqa} \& \squadcolor{SQuAD} \& \hotpotcolor{HotpotQA} & \underline{98.2} & \underline{98.3} & \underline{98.2} & 89.3 & \underline{82.4} & \underline{85.8} & \textbf{94.5} & \textbf{92.7} & \textbf{93.6} & \textbf{85.5} & \textbf{71.0} & \textbf{77.6} \\
\bottomrule
\end{tabular}
}
\caption{
Performance comparison of models fine-tuned on synthetic versus gold in-domain data. Our \synqa approach demonstrates superior generalization while maintaining competitive performance on in-domain tasks. \textbf{Bold} indicates the best-performing method; \underline{underlined} indicates our method when it ranks second. 1P = models trained with a single pass over training data; 1M = models trained with 1M samples (matched to \synqa data size).
}
\label{table:fine-tuned-models}
\end{table*}

In Table~\ref{table:fine-tuned-models}, we compare models trained on synthetic and gold in-domain context attribution datasets. As expected, fine-tuning on in-domain gold datasets (SQuAD and HotpotQA) yields highly specialized models that perform well on in-domain data. However, models trained on data obtained by \synqa exhibit competitive performance on in-domain tasks and consistently surpass in-domain-trained models on out-of-domain datasets.
Fine-tuning models on in-domain data (SQuAD and HotpotQA, in addition to \synqa) further improves recall, particularly on HotpotQA. Specifically, recall improves from 69.4\% to 82.4\% when adding in-domain data to models trained with \synqa. We observe a similar trend for QuAC-ST and CoQA-ST: \synqa-trained models already exhibit high precision, but recall is further enhanced when adding in-domain data. This suggests that while \synqa effectively teaches the model to precisely attribute evidence, fine-tuning with domain-specific data allows it to capture a broader set of relevant evidence. Such out-of-domain generalization is crucial for practical deployments, where models must handle diverse, unseen contexts that often differ substantially from their training data.


\subsubsection{Comparison to Zero-Shot and Fine-Tuned Models in Dialogue Settings}\label{sec:experiments-dialog}
\begin{table*}[t]
\centering
\resizebox{1.0\textwidth}{!}{
\begin{tabular}{lcccccccccccccc} 
\toprule

\multirow{3}{*}{Model} & \multirow{3}{*}{Training data} 

& \multicolumn{12}{c}{\textbf{Out-of-Domain}} \\ \cmidrule(lr){3-14}

& & \multicolumn{3}{c}{\quaccolor{QuAC}} & \multicolumn{3}{c}{\coqacolor{CoQA}} 
& \multicolumn{3}{c}{\orquaccolor{OR-QuAC}} & \multicolumn{3}{c}{\doqacolor{DoQA}} \\ 

\cmidrule(lr){3-5} \cmidrule(lr){6-8} \cmidrule(lr){9-11} \cmidrule(lr){12-14}

 &  & P & R & F1 & P & R & F1 & P & R & F1 & P & R & F1 \\ 
\midrule
\textbf{\textit{Baselines}} \\
Llama | 1B & Zero-shot & 30.8 & 45.5 & 36.8 & 39.4 & 37.9 & 38.6 & 33.0 & 46.6 & 38.6 & 12.2 & 22.6 & 15.9 \\
Mistral | 7B & Zero-shot & 76.6 & 81.8 & 79.1 & 67.6 & 61.3 & 64.3 & 82.5 & 85.1 & 83.8 & 74.9 & 77.9 & 76.4 \\
Llama | 8B & Zero-shot & 84.7 & 88.8 & 86.7 & 79.3 & 72.0 & 75.5 & 88.0 & 91.3 & 89.6 & 77.9 & 91.4 & 84.1 \\
Mistral-NeMo | 12B & Zero-shot & 85.7 & 85.4 & 85.5 & 81.9 & 68.4 & 74.5 & 88.9 & 88.8 & 88.8 & 86.0 & 84.2 & 85.1 \\
Llama | 70B & Zero-shot & 88.5 & 87.7 & 88.1 & \textbf{88.3} & \textbf{74.9} & \textbf{81.1} & 81.7 & 86.3 & 83.9 & 85.2 & 82.0 & 83.5 \\
\midrule
\textbf{\textit{Baselines}} \\
Llama | 1B & \squadcolor{SQuAD} \& \hotpotcolor{HotpotQA} (1P) & 71.3 & 66.8 & 69.0 & 79.0 & 64.2 & 70.8 & 61.6 & 57.5 & 59.5 & 67.4 & 57.8 & 62.2 \\
Llama | 1B & \squadcolor{SQuAD} \& \hotpotcolor{HotpotQA} (1M) & 52.6 & 49.3 & 50.9 & 61.2 & 50.2 & 55.2 & 48.5 & 44.6 & 46.5 & 53.2 & 49.1 & 51.1 \\ \midrule
\textbf{\textit{Ours}} \\
Llama | 1B & \syntheticcolor{\synqa} & \textbf{91.3} & \underline{91.4} & \underline{91.3} & 81.7 & 71.4 & 76.2 & \textbf{92.6} & \underline{95.3} & \textbf{94.0} & \textbf{86.3} & \underline{94.5} & \textbf{90.2} \\
Llama | 1B & \syntheticcolor{\synqa} \& \squadcolor{SQuAD} \& \hotpotcolor{HotpotQA} & \underline{91.1} & \textbf{92.2} & \textbf{91.7} & \underline{82.3} & \underline{73.2} & \underline{77.5} & \underline{90.3} & \textbf{96.4} & \underline{93.2} & \underline{85.1} & \textbf{96.0} & \textbf{90.2} \\
\bottomrule
\end{tabular}
}
\caption{Context attribution on QuAC, CoQA, OR-Quac, and DoQA (dialogue data); all datasets are out-of-domain (i.e., we do not use the training sets). Our \synqa models outperform fine-tuned and larger zero-shot models,  despite their size advantage. \textbf{Bold} denotes best method, \underline{underline} our method when second best. 1P: models trained with a single pass over the training data. 1M: models trained with 1M samples (matched to \synqa data size).}
\label{table:dialog-datasets}
\end{table*}

We evaluate dialogue context attribution, for which we do not use any gold in-domain training data (Table~\ref{table:dialog-datasets}).\footnote{Note that \synqa contains dialogue-specific (i.e., multi-turn) training data. See \S\ref{sec:synqa_method} and Appendix~\ref{app:qa-generation} for details.} Here, models must handle follow-up questions that rely on previous turns, often involving coreferences and other dialogue-specific complexities. Since all dialogue datasets can be considered out-of-domain compared to SQuAD and HotpotQA (dialogue vs. single-turn; different sources—e.g., DoQA is derived from StackExchange while SQuAD and HotpotQA are derived from Wikipedia), the patterns we observe differ in some cases from the single-turn results.
As expected, zero-shot LLMs exhibit a strong size-performance correlation, with larger models consistently outperforming smaller ones---even those fine-tuned on single-turn question-answer attribution (trained on gold SQuAD and HotpotQA). However, fine-tuning smaller models with \synqa data leads to superior performance, surpassing both their fine-tuned counterparts and much larger zero-shot LMs.
Unlike Table~\ref{table:fine-tuned-models}, training on SQuAD and HotpotQA does not lead to consistent improvement in dialogue settings. Since these datasets consist of single-turn QA, they are less beneficial for improving performance on multi-turn QA data.
As with the single-turn results, zero-shot LLMs already exhibit high precision, but models using \synqa data either match or surpass them while improving recall, further suggesting that \synqa enhances the model's ability to extract a more complete set of supporting evidence, even in dialogue settings.


\subsubsection{Scaling Trends and Generalization Performance}\label{sec:scalling-trends}

\begin{figure*}[t]
    \centering
    \begin{subfigure}{0.48\linewidth}
        \centering
        \includegraphics[width=\linewidth]{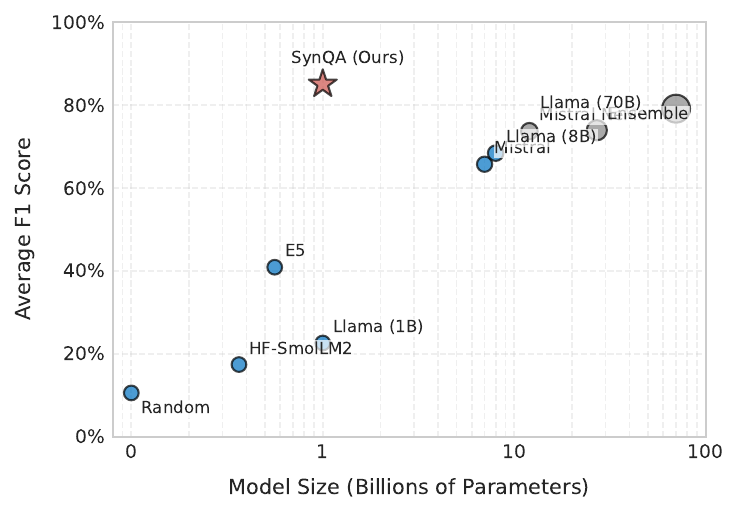}
        \caption{Model performance vs.~size.}
        \label{fig:size_performance}
    \end{subfigure}
    \hfill
    \begin{subfigure}{0.48\linewidth}
        \centering
        \includegraphics[width=\linewidth]{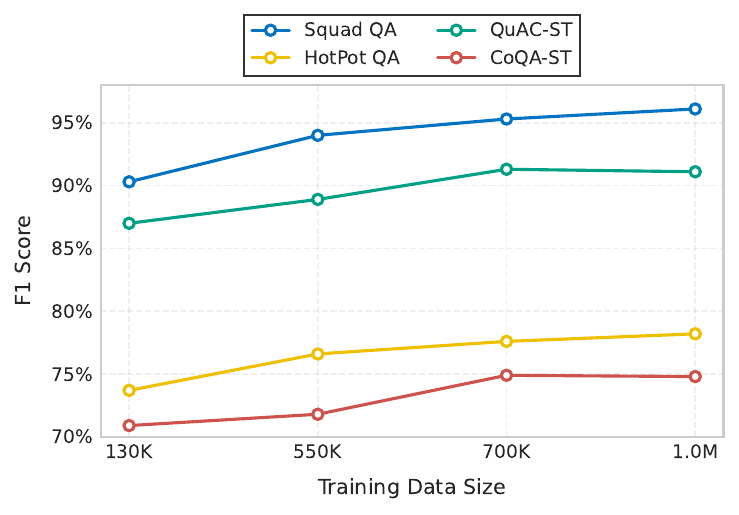}
        \caption{F1 score vs.~training data size.}
        \label{fig:data_quantity}
    \end{subfigure}
    \caption{Comparison of model performance and scalability. (a) Larger zero-shot models achieve good F1 scores, but our method \synqa (based on Llama 1B) outperforms them while being orders of magnitude smaller. (b) Performance improves consistently with more \synqa training data, highlighting its scalability.}
    \label{fig:combined}
\end{figure*}

Fig.~\ref{fig:size_performance} shows F1 scores averaged across datasets, with model size on the x-axis and performance on the y-axis. Models trained on \synqa-generated data significantly outperform their baseline zero-shot counterparts, while also achieving superior performance compared to zero-shot LLMs that are orders of magnitude larger. This shows our method is highly data-efficient, enabling small models to close the gap with much larger counterparts.

In Figure~\ref{fig:data_quantity}, we analyze model performance as the quantity of synthetic training data increases, reporting F1 scores separately for in-domain and out-of-domain datasets. As we scale data quantity, performance improves consistently across datasets for single-turn context attribution. This trend highlights the scalability of our approach, indicating that further gains can be achieved by increasing the quantity of the synthetic data.

\subsubsection{User Study: \synqa increases efficiency and accuracy assessment}\label{sec:user-study}
We conducted a user study to evaluate the efficiency and accuracy of verifying the correctness of LLM-generated answers using context attribution. Our hypothesis is that higher-quality context attributions, visualized to guide users, facilitate faster and more accurate verification of LLM outputs. Specifically, in each trial, we presented users with a question, a generated answer, and relevant context, along with
attributions visualized as highlights. Their task was to leverage these attributions to judge if the answer was correct w.r.t.~a provided context. See Figure~\ref{fig:user_interface} in Appendix~\ref{app:user_study}.

The study compares three scenarios:
\begin{inparaenum}[(i)]
\item 
\textbf{No Alignment:} a baseline condition without context attributions, requiring users to manually read and verify the answer against the entire context;
\item 
\textbf{Llama 1B (Zero-shot):} context attributions generated by the Llama 1B model were visualized;\footnote{We specifically did not choose a large model as we would want to run these models in real time in real user applications.}
\item 
\textbf{\synqa}: context attributions generated by our approach were visualized.
\end{inparaenum}

We employed a within-subjects experimental design for our human evaluation (with 12 participants), ensuring that the same participants evaluate all the aforementioned alignment scenarios, thus requiring fewer participants for reliable results \cite{greenwald:1976}. However, this can be susceptible to learning effects where participants perform better in later scenarios, because they learned the task from previous examples. To mitigate this, we counterbalanced the scenario order using a Latin Square design \cite{belz:2010,bradley:1958}, where each alignment scenario appears in each position an equal number of times across all participants. Finally, we randomized the example order within each scenario per participant. For each example, we measured: \textbf{verification time} (seconds from display to judgment submission) and \textbf{verification accuracy} (binary correct/incorrect judgment).

\begin{figure}[t]
    \centering
    \includegraphics[width=1.0\linewidth]{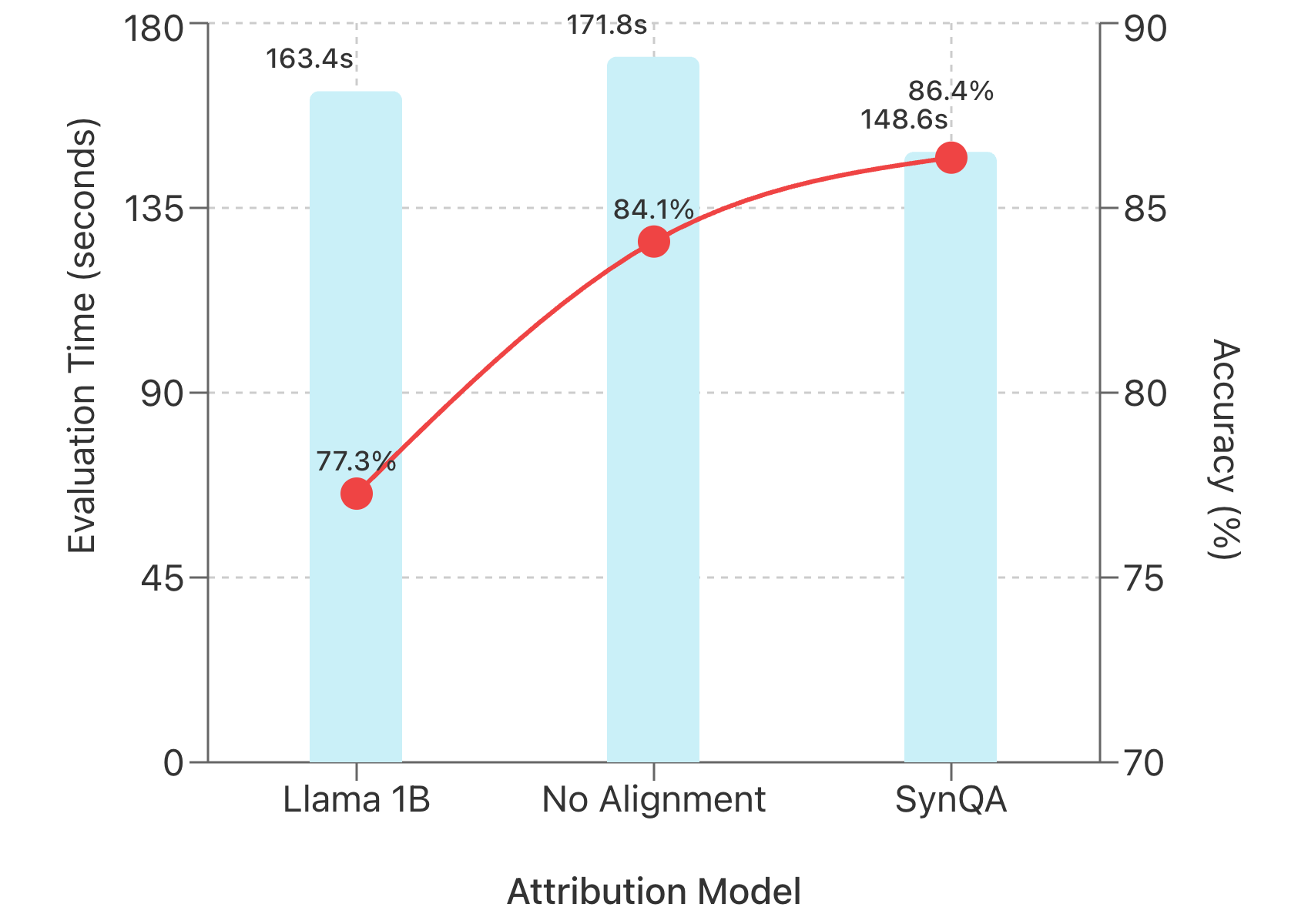}
    \caption{Relationship between Evaluation Time (seconds) and Accuracy (\%) for three answer verification settings:  \emph{Llama 1B (Zero-shot)}, \emph{No Alignment} and \synqa. \synqa demonstrates the lowest evaluation time and highest accuracy, indicating its superior performance in facilitating efficient and accurate answer verification.}
    \label{fig:user_study}
\end{figure}

\noindent \textbf{Results.} We observed a clear trend in verification performance across the different attribution settings, with \synqa demonstrating superior effectiveness (Fig.~\ref{fig:user_interface}). \synqa has the lowest average verification time per example (\textbf{148.6} seconds), significantly faster than \emph{No Alignment} (171.8 seconds) and attributions from \emph{Llama 1B} (163.4 seconds). Concurrently, in terms of verification accuracy, \synqa achieved the highest average accuracy (\textbf{86.4\%}). While \emph{No Alignment} (84.1\%) and \emph{Llama 1B (77.3\%)} also yielded reasonable accuracy, attributions from \synqa are clearly of higher quality helping users be more accurate.

\section{Related Work}

Work on context attribution for QA can be split into two categories: (1) in-line citation generation: LLMs are instructed to generate citations along with the generated answer; (2) post-hoc context attribution: perform the attribution \emph{after} the LLM generates the answer. In the following, we outline these works and their differences from our work. For a more comprehensive discussion on related work, see Appendix~\ref{app:rel_work}.

\subsection{In-line Citation Generation}

In this setup, researchers use LLMs to produce in-line citations along with the generated text \cite{bohnet2022attributed,gao2023enabling,Huang2024AdvancingLL}. This typically works on paragraph or document level. One line of work focuses on fine-tuning methods for tackling the problem \cite{gao2023enabling,Schimanski2024TowardsFA,Berchansky2024CoTARCA,patel2024towards}, while another line of work proposes synthetic data generation methods for fine-tuning such models \cite{Huang2024LearningFG,Huang2024AdvancingLL}. 
\citet{slobodkin2024attribute} propose a fine-grained task, where the attributions are on sentence level, because such granularity is more useful to human end users. Since generating such in-line citations can result in producing completely made-up citations, \citet{Yue2023AutomaticEO} propose a task that checks whether in-line generated citations from LLMs are attributable or not. 
Unlike such approaches, we focus on post-hoc context attributions, because this directly predicts a link to a factual source, and therefore avoiding the risk of making up the source.

\subsection{Post-hoc Context Attribution}

In post-hoc context attribution, the aim is to determine which parts of the context are attributable to an already answered question \cite{Yang2018HotpotQAAD}. There has been a significant amount of work on training models for the context attribution problem on sentence-level for multi-hop QA \cite{zhang2024end, ho2023analyzing,yin2023rethinking,fu2021decomposing,tu2020select,fang-etal-2020-hierarchical}. However, they do not investigate this problem in the context of LLMs. Moreover, the methods are constrained \emph{only} to multi-hop QA, and are not tested on broader QA context, such as dialogue QA. In our work, we propose methods that use LLMs as data generators. This allows us to better generalize and cover multiple QA settings simultaneously, therefore better matching real-world needs. 

Another line of work focuses on coarse-level granularity and provides attributions either on paragraph level \cite{rashkin2023measuring,Menick2022TeachingLM} or document level \cite{Nakano2021WebGPTBQ,Gao2023RARRRA,Buchmann2024AttributeOA}. However, in a user study \citet{slobodkin2024attribute} observe that such granularity level is not optimal for humans when manually fact-checking LLM-generated content. Their experiments suggest that sentence-level granularity is ideal for humans. This is why we adopt sentence-level granularity in our work, despite this being a harder task. On the other hand, there has been work that focuses on the other extreme: assigning context attributions on sub-sentence level \cite{CohenWang2024ContextCiteAM,Phukan2024PeeringIT, Ramu2024EnhancingPA}. While these approaches provide more granular attributions, they are computationally more expensive, which hinders their practical usability. Our work ensures that models can be run in real-time to make them practical for end users.

\section{Conclusion}
We investigated the task of context attribution in QA. We focused on approaches that enhance attribution performance without relying on prohibitive human annotations. Our proposed data synthesis strategy, \synqa, enables the generation of high-quality synthetic attribution data, leading to substantial improvements in fine-tuned small models.

Through extensive experiments on six datasets across single turn QA and dialogue QA attribution, we demonstrated that small models fine-tuned with \synqa data (i) significantly outperform models trained on alternative synthetic attributions, (ii) exceed the performance of zero-shot LLMs that are orders of magnitude larger, and (iii) generalize better to out-of-domain distributions compared to models trained on gold in-domain data. These findings suggest that \synqa reduces reliance on large-scale human-labeled datasets, while improving attribution robustness across diverse scenarios.


Finally, our user study validates the practical utility of fine-tuned small models in real-world question-answering applications. These results highlight the viability of scalable, data-efficient context attribution techniques, paving the way for more interpretable and trustworthy AI systems. 




\section*{Limitations}
While our work demonstrates the effectiveness of \synqa for context attribution in question answering, we leave some important directions for future research. First, all models we train operate exclusively at the sentence level. Even though \citet{slobodkin2024attribute}, through a user study, found that sentence-level granularity of context attribution QA is probably the best-suited granularity for manual verification of LLM output, it might not always be optimal granularity for attribution in other tasks. Namely, some context elements might be better captured at different levels: e.g., from individual phrases to multi-sentence passages—depending on the semantic structure of the text.

Second, while we evaluated our approach on OR-QuAC, we have not fully explored context attribution in retrieval-augmented generation (RAG) settings with dialogue. This presents unique challenges since models must simultaneously handle dialogue-style questions and continuously evolving context. Future work should examine how attribution models can adapt when relevant context changes throughout a conversation.

Third, we focused primarily on question answering, however, context attribution is valuable for many other natural language processing tasks, e.g., in text summarization, attributing summary sentences to source document segments could enhance transparency and fact-checking capabilities. Future research should examine how \synqa's synthetic data generation approach can be adapted for different tasks, potentially revealing task-specific challenges and nuances.

Fourth, our user study (\S\ref{sec:user-study}), while providing valuable initial insights into the effectiveness of context attribution to help users verify the LLM model outputs in QA settings, was conducted with a limited sample of 12 participants. A larger-scale study with more participants would strengthen the statistical validity of our findings and potentially reveal more nuanced patterns. Future work should extend this evaluation to a more diverse and larger participant pool, ideally, including users with varying levels of domain expertise and familiarity with language model outputs.
\section*{Acknowledgements}
We thank Andreas Ripke for his support with hosting the LLMs used in our experiments. We also thank Dina Trajkovska for turning our messy ideas into beautiful figures that actually make sense.

\clearpage

\bibliography{custom}

\clearpage

\appendix
\section{Comprehensive Discussion on Related Work}
\label{app:rel_work}

We split the related work papers into several categories: tasks, datasets, methods and metrics.

\subsection{Context Attribution Tasks} 

\paragraph{Attributable to Identified Sources (AIS):} given a generative text $t_g$ and a context text $t_c$, is $t_g$ attributable to $t_c$? \citet{rashkin2023measuring} propose a manual framework that defines the AIS task and evaluates the AIS scores across several NLP tasks, namely conversational QA \cite{Anantha2020OpenDomainQA,Dinan2018WizardOW}, text summarization \cite{Nallapati2016AbstractiveTS} and table-to-text \cite{Parikh2020ToTToAC}. Our work differentiates in three important aspects: (1) we focus on a broader QA setup (i.e., single-question QA and conversational QA), which makes our work a subset of the broader AIS task; (2) we focus on more fine-grained level: our attributions are not on the entire text level, but rather on a sentence level, which has been shown in user studies to be more useful to end-users \cite{slobodkin2024attribute}; (3) the AIS task entails a manual evaluation framework, while our work provides automatic evaluation with golden data.


\paragraph{In-line Citation Generation:} 
uses LLMs to produce in-line citations along with the generated text \cite{bohnet2022attributed,gao2023enabling,Huang2024AdvancingLL}. This typically works on paragraph or document level. 
\citet{slobodkin2024attribute} propose a fine-grained task, where the attributions are on sentence level, because such granularity is more useful to human end users. Because generating such in-line citations can result in producing completely made up citations, \citet{Yue2023AutomaticEO} propose a task that checks whether the in-line generated citations from LLMs are actually attributable or not. Instead of using binary attributable/non-attributable labels (like with AIS), they propose more fine-grained labels for this problem: attributable, extrapolatory, contradictory and non-attributable. Contrary to such approaches, our work focuses on post-hoc context attributions: given an answer to a question, find the sentences in the context that support the factuality of the answer.


\paragraph{Post-hoc Attribution:} determines which parts of the context are attributable to an already answered question \cite{Yang2018HotpotQAAD}. Within the post-hoc context, there are two other subcategorizations of the task: \textit{contributive} and \textit{corroborative post-hoc attribution} \cite{CohenWang2024ContextCiteAM}.

\paragraph{Post-hoc Attribution (Contributive):} ContextCite \cite{CohenWang2024ContextCiteAM} and Mirage \cite{Qi2024ModelIA} define a post-hoc task that aims at discovering which parts of the context \textit{caused} the LLM to generate the particular response. Their evaluation methods, however, are based on  proxy metrics that do not rely on golden annotations, while in our work we rely on automatic annotations that rely on golden data. 

\paragraph{Post-hoc Attribution (Corroborative):} this task is similar to contributive post-hoc attribution. The difference is that the constraint for causality is not necessarily enforced, but should support the factuality of the statement \cite{CohenWang2024ContextCiteAM}. Many works are based on coarse-grained level and provide attributions on either paragraph level \cite{Menick2022TeachingLM}, document level \cite{Nakano2021WebGPTBQ} or on multi-document level, where they have a RAG component that retrieves the documents that are potentially attributable \cite{Gao2023RARRRA,Buchmann2024AttributeOA}.



\paragraph{Context Attributions to other Modalities.} Other line of work maps the attributions to other modalities, such as knowledge graphs \citet{Dammu2024ClaimVerEC}. Similarly, \citet{Maheshwari2024PresentationsAN} take multi document collection as input, construct a graph of narratives, and then generate a presentation (i.e., slides) for the topic, along with attributions from the generated content of the slides with the original documents. We do not investigate such cases, and focus on attributing answers to sentences within the user-provided context.

\paragraph{Post-hoc Attribution for Text Summarization.}
\citet{ernst2021summary} proposed a task, dataset and baseline model (dubbed SuperPAL) for detecting attributions for text summarization. In a followup work, \citet{ernst2022proposition} an extension of the task, this time for clustering propositions for text summarization and \citet{ernst2024power} extend this to multi-document summarization. \citet{krishna2023longeval} investigate whether such text summarization alignments are helpful for humans. In our work, we focus on the question answering (QA) task.

\subsection{Datasets}

\paragraph{AIS.} \citet{rashkin2023measuring} proposed the AIS dataset, which contains three tasks: question answering, table-to-text and text summarization. Here, for each data point, there is a query and an LLM-generated response, along with label by humans whether it is fully attributable or not. This data is on paragraph and document level, and lacks the granularity of a sentence level. Therefore, we do not use it in our work.

\paragraph{HotpotQA.} With HotpotQA \cite{Yang2018HotpotQAAD}, the authors propose an explainable multi-hop QA dataset. The dataset also contains attribution links (i.e., explanations) for the answers: spans of text that belong to the input context, which are supporting the statement in the answer. The authors set up baselines for measuring the ability of attributions of models on sentence level, which is in line with what we do. In our work, we integrated HotpotQA as part of our setup for both training and testing.

\paragraph{AttributionBench.} This is a benchmark for attribution evaluation of LLM generated content \cite{Li2024AttributionBenchHH}. In particular, the benchmark assesses whether the assigned attribution on a generated text is actually attributable. In particular, given a query, response set $\mathcal{R}$ (containing claims) and evidence set $E$, the task is to label as "attributable" or "not attributable" every claim against $E$. This work operates on a coarse-grained level (paragraphs or whole documents). Similarly, \citet{Yue2023AutomaticEO} proposed another dataset for evaluating attribution of LLM-generated text, same on paragraph level. In our work, we focus on sentence-level context attribution. Therefore, we did not include this dataset in our work.

\paragraph{Conversational QA.} CoQA \cite{Reddy2018CoQAAC} is a conversational QA dataset, which contains a context, questions-answer pairs between two people (teacher and student) and sentence-level supporting evidence for the context. We use this dataset in our evaluation to test the out-of-domain capabilities of LLMs for context attribution. We also use QuAC \cite{Choi2018QuACQA} and ORConvQA \cite{qu2020open}, which are conversational QA datasets similar to CoQA. 

\paragraph{QASPER.} This dataset is from the scientific domain \cite{dasigi2021dataset}. The dataset contains title and an abstract of a paper, question and answer about the content. This data has information on paragraph level, not on sentence level. Therefore, we do not use it in our experiments.

\paragraph{WikiQA.} \citet{Dammu2024ClaimVerEC} use WikiQA \cite{yang2015wikiqa}, because it's Wikipedia-based dataset, which can be linked to Wikipedia-derived KG like Wikidata \cite{vrandevcic2014wikidata}. In our work, we focus only on text modality, which is why we do not include this dataset into our evaluation.

\paragraph{WikiNLP.} The WikiNLP dataset \cite{gashteovski2019opiec} is a dataset that contains the entire English Wikipedia, along with linguistic annotations (e.g., POS tags, dependency parse trees, etc.) and semantic annotations (e.g., NER tags and entity links). We use this dataset for the synthetic data generation, because it keeps the linked information from the Wikipedia articles, which are annotated by humans. For this reason, the dataset has been used in wide range of tasks in research, mostly for information extraction \cite{dukic-etal-2024-leveraging,kotnis2023open,kotnis2022,gashteovski2020aligning}, but also for other tasks such as clustering \cite{viswanathan-etal-2024-large}, open link prediction \cite{broscheit-etal-2020-predict} and entity linking \cite{nanni2019eal,radevski-etal-2023-linking}


\subsection{Metrics and Evaluation} 

\paragraph{AIS.} The AIS framework \cite{rashkin2023measuring} is human annotation framework. Given a generated text chunk and a context chunk (this can be sentence, paragraph or document), a human evaluated whether the generated text chunk is fully attributible or not. It is basically a binary classification problem. In their data, the authors focus on document level granularity, which is not useful for humans. In our setup, we check for each sentence in the context if it supports the answer.

\paragraph{AutoAIS.} To evaluate the attributed information, \citet{slobodkin2024attribute} use \textbf{AutoAIS metric}: an NLI-based scoring. Prior studies have shown that this metric highly correlated with human annotations \cite{bohnet2022attributed,gao2023enabling}. This is an extension to the AIS metric. We do not use proxy metrics, but rely on golden annotations by humans.

\paragraph{AttrScore.} With AttrScore, \citet{Yue2023AutomaticEO} consider LLM generated content on the one hand and citation documents on the other hand. Then, the score evaluates whether a provided citation is attributable, extrapolatory, contradictory or non-attributable. Essentially, it is an extension of AIS, such that it provides more fine grained labels for the provided citations. The AttrScore is basically a fine-tuned LLM that provides these scores.

\paragraph{Unsupervised Metrics.} ContextCite proposed the Top-k-drop and LDS metric to evaluate the causal post-hoc attribution. These metrics do not require labeled data. \citet{Berchansky2024CoTARCA} uses ROUGE and BERTScore to evaluate their results. We do not use unsupervised metrics and rely on automated evaluation with golden annotations by humans.

\subsection{Methods} 

\paragraph{Multihop QA.} There has been significant amount of work on tackling the context attribution problem on sentence level for multi-hop QA \cite{zhang2024end, ho2023analyzing,yin2023rethinking,fu2021decomposing,tu2020select,fang-etal-2020-hierarchical}. While we also investigate this problem, in contrast to our work, these works focus \emph{only} on the multihop QA task. In our work, we also explore other QA setups, including conversational QA with different domains. Moreover, these papers do not investigate the capabilities of LLMs regarding the context attribution problem, but rather propose specific methods that are tailor-made for the multihop QA problem, which involves both answering the questions and providing supporting sentences to the answers. 

\paragraph{In-line Citation Generation.} Another line of work focuses on guiding LLMs to generate in-line citations along with the generated text \cite{Li2023ASO}. \citet{slobodkin2024attribute} tackle this problem on a sentence level, but do not investigate the post-hoc context attribution case. Moreover, they rely on proxy metrics such as AutoAIS \cite{Gao2023RARRRA} and BERTScore \cite{zhang2020bertscore}. Similarly, \citet{bohnet2022attributed} proposes methods for in-line citation generation, but this work is more coarse-grained and focuses on paragraph and document level. They also report their findings on proxy metrics. Their method is based on retrieval and they do not investigate the LLMs' capabilities thoroughly. \citet{gao2023enabling} assign citations to LLM-generated content, where they retrieve the information from a large collection of documents (also, it's on paragraph and document level, not on sentence level). START \cite{Huang2024AdvancingLL} propose a data synthetic generation method for in-line citation generation on document level, where each citation refers to an entire document. FRONT \cite{Huang2024LearningFG} also investigates synthetic data generation of in-line citation generation, where the citations assigned to the sentences in the output are entire documents. Similarly, \citet{Schimanski2024TowardsFA} propose a synthetic data generation pipeline for fine-tuning models that solve the same problem. 
\citet{Berchansky2024CoTARCA} use Chain-of-Thought approaches and fine-tuning smaller LLMs in order to solve this problem. \citet{patel2024towards} also fine-tune a model specifically for this task, and the attributions are on paragraph level.

\paragraph{Post-hoc Context Attribution.} \citet{Ramu2024EnhancingPA} propose template-based in-context learning method for post-hoc context attribution. In particular, they use standard retrievers as a first step to pre-rank the text (e.g., BM25 and dual encoders \cite{ni2022large}) and then they use LLMs to classify (i.e., rerank) the relevant sentences. 
ContextCite \cite{CohenWang2024ContextCiteAM} uses ablation-based methods to infer the attributions of post-hoc generated text. 

\paragraph{User Study.} Recent work has called for a more human-centric research in NLP \cite{kotnis2022human}. In such work, the idea is to involve the user (i.e., the final stakeholder) in the process of research, which is typically done with some forms of user studies \cite{rim2024human,xu-etal-2024-human,ilievski2024aligning}. In this spirit, we want to verify whether our models are useful for end users. To this end, we performed a user study, whereas users were asked to solve a fact-checking task with the use of our context attribution model. We found that our approach does indeed make human end-users faster in performing the manual fact-checking task (for details of our user study, see Section~\ref{sec:user-study}).

\section{Method for Synthetic Data Generation}
\label{app:synthetic_data}

\subsection{Multi-hop Generation of Attribution Data}\label{app:data-collection}
To generate synthetic data with the use of Wikipedia, we use the WikiNLP dataset \cite{gashteovski2019opiec}. It contains the text from all Wikipedia articles along with annotations for links within the text that link to other Wikipedia articles. The main idea is to use the links in order to imitate reasoning hops across different (related) articles. Therefore, we filter out all articles that either do not contain links or that contain links to articles that do not contain links. Finally, for each article, we use only the first paragraph, because this is considered to be the paragraph that contains the most ``definitional information'' \cite{bovi2015}; i.e., information that precisely describes the target concept of the article and contains the most important information about it.

Then, for each article, we randomly select a sentence that contains at least one link to another Wikipedia article.\footnote{To make sure we have multi-hop scenario, we also check if the other Wikipedia article also contains at least one valid link to another Wikipedia article.} Each of the sentences that we sample serves as ground truth for the context attribution. With these sentences, we then prompt an LLM to generate a question-answer pair.

\subsection{Question-Answer Pairs Generation}\label{app:qa-generation}
To generate context attribution training data that requires multi-hop reasoning, using the multi-hop chain of sentences, we prompt the LLM (see \S\ref{app:prompts-multi-hop} for the prompt we use) by providing it \textit{only} the formed chain as the ground truth attributions, which the LLM must use for the question-answer pair generation. The LLM generates a question-answer pair that can only be answered using the information in these supporting sentences, ensuring the pairs are grounded in the provided evidence.

For dialogue-centric context attribution data, we simply provide the LLM with a single Wikipedia article, and prompt the model to generate multi-turn QA pairs and provide the attribution sentence for each QA pair (see \S\ref{app:prompts-dialogue} for the prompt we use).

\subsection{Distractors Mining}\label{app:distractors-mining}
In realistic scenarios, whether the context is user-provided or retrieved through RAG, the system typically encounters multiple context documents that are highly similar to those containing the evidence sentences. To bridge this gap between our synthetically generated training data using \synqa and the data models encounter ``in the wild'', we augment each training sample with hard negative distractor articles. We obtain embeddings using E5 \cite{wang2022text} for each Wikipedia article in our collection. Then, for each article containing a supporting sentence for the question-answer pair, we randomly sample up to three distractor articles that share semantic similarity with the ground truth article. This process increases the difficulty of the training data, producing models better equipped to handle diverse testing scenarios.

\subsection{Comparison to HotpotQA}

Although our method is inspired by HotpotQA \cite{Yang2018HotpotQAAD}, note that we do not aim to recreate the HotpotQA dataset. Our method has significant differences, enabling us to generate as much data as needed, with a higher domain variability. 

Particularly, the method of \citet{Yang2018HotpotQAAD} is more curated and has a human in the loop in multiple steps. First, the authors manually select the target entities (and, with that, the target articles from which the annotators create the question and answer pairs). The reason for this is that many highly specialized articles---e.g., the article for IPv4 protocol---are not suitable for crowd-workers to both identify meaningful questions and provide answers for those questions. Our approach does not have this constraint and, therefore, produces data that has much higher domain variability. 

Second, their method uses Amazon Mechanical Turk workers to annotate the questions, answers, and attribution sentences. In our case, we automatically select the hopped sentences (which serve as gold context attribution data), and then we use these sentences to generate question-answer pairs with an LLM.

Third, while HotpotQA always enforces multi-hop QA pairs, we do not instruct the LLM to do that. Rather, we first allow the LLM to decide whether generating such a multihop QA pair is possible for the incoming context attribution sentences. If so, then the LLM generates multihop QA pairs. Otherwise, it generates direct QA pairs that do not need hops; i.e., QA pairs like in SQuAD \cite{Rajpurkar2016SQuAD1Q}.

Fourth, the HotpotQA annotation method does not allow for dialogue QA. In our method, we also create dialogue context attribution data.

With these differences in mind, we showed that, compared to HotpotQA, our data generation method exhibits the following advantages: (1) generates data with higher domain variability; (2) goes beyond multi-hop QA and also generates direct QA pairs (like SQuAD) as well as dialogue QA data; (3) the data is generated in completely automatic manner without the involvement of humans.


\section{Multi-turn to single-turn conversations}\label{app:multi_to_single}
We convert the multi-turn Quac and CoQA datasets to single-turn (Quac-ST and CoQA-ST) using Llama 70B. In Table~\ref{table:qa-rephrasing}, we provide examples of the questions and answers before and after converting them.

\begin{table*}[t]
\centering
\resizebox{0.9\textwidth}{!}{
\begin{tabular}{llp{0.35\textwidth}p{0.35\textwidth}} 
\toprule
\textbf{ID} & \textbf{Type} & \textbf{Original} & \textbf{Rephrased} \\
\midrule
Q1 & Question & What happened in 1983? & What significant event occurred in Anna Vissi's life in 1983? \\
A1 & Answer   & In May 1983, she married Nikos Karvelas. & In May 1983, Anna Vissi married Nikos Karvelas, a composer. \\
\midrule
Q2 & Question & Did they have any children? & Does Anna Vissi have any children? \\
A2 & Answer   & She gave birth to her daughter Sofia. & Anna Vissi gave birth to her daughter Sofia in November. \\
\midrule
Q3 & Question & What collaborations did she do with Nikos? & What is the nature of Anna Vissi's collaborations with Nikos Karvelas? \\
A3 & Answer   & After their marriage, she started a close collaboration with Karvelas. & Since 1975, Nikos Karvelas has composed songs for all of Anna Vissi's releases, which have become gold or platinum. \\
\bottomrule
\end{tabular}
}
\caption{Examples of the original (Quac and CoQA) and rephrased questions and answers (Quac-ST and CoQA-ST).}
\label{table:qa-rephrasing}
\end{table*}
\section{Prompts to generate \synqa synthetic training data}\label{app:prompts}

\subsection{Multi-hop reasoning context attributon}\label{app:prompts-multi-hop}

\begin{prompt}
\textbf{SYSTEM PROMPT}

You are tasked with generating a concise and focused question-answer pair using information from provided Wikipedia sentences. Follow these instructions carefully:

1. You will be provided with multiple Wikipedia articles, each containing:

   - The title of the article.
   
   - One specific sentence from the article.

2. Your goal is to generate a \textbf{short, factual question} and a \textbf{concise answer}, ensuring:

   - The question-answer pair is grounded in the provided sentences.
   
   - The reasoning is logical, clear, and references all sentences used.

3. \textbf{Key Constraints}:

   - Questions must address a \textbf{single coherent topic} or concept that can be logically inferred from the provided sentences.
     
   - Avoid combining unrelated pieces of information into a single question.
   
   - The \texttt{"reasoning"} must explain how each sentence in the \texttt{"ids"} field contributes  to answering the question but should remain \textbf{brief} and \textbf{to the point}.

4. Aim for \textbf{brevity}:

   - Questions should be concise and avoid unnecessary details.
   
   - Answers should be short, typically no more than one sentence.
   
   - Keep the reasoning concise, focusing only on the necessary logical connections.

5. Multi-hop reasoning is encouraged but must be natural and focused:

   - Combine information only when it is logical and directly relevant to the question.
   
   - Do not create overly complex questions that combine weakly related information.

6. Provide your response in \textbf{raw JSON format} with the following keys:

   - \texttt{"question"}: A concise and clear question string.
   
   - \texttt{"answer"}: A short and factual answer string.
   
   - \texttt{"ids"}: A list of JSON-compatible arrays (e.g., \texttt{[[0, 0], [1, 0]]}) representing the indices of all sentences used to generate the question-answer pair.
   
   - \texttt{"reasoning"}: A brief explanation of how \textbf{each sentence in \texttt{"ids"}} was used to 
     generate the question-answer pair.

\textbf{Important Notes}:

- Ensure the question-answer pair is entirely self-contained and logically consistent.

- Do not include unnecessary or weakly related information in the question or answer.

- Avoid introducing information not present in the provided sentences.

- Do not include additional formatting, explanations, or markdown in your response.
\end{prompt}

\begin{prompt}
\textbf{USER PROMPT}

Here are the titles and sentences:

Title: [First Article Title]

[0, 0] [First sentence from the article]

Title: [Second Article Title]

[1, 0] [Second sentence from the article]

Title: [Third Article Title]

[2, 0] [Third sentence from the article]

Use the provided sentences to generate a question-answer pair following the specified guidelines. Respond \textbf{only in raw JSON} with no additional formatting or markdown.
\end{prompt}

Given the \textbf{SYSTEM} and the \textbf{USER} prompt, the LLM is generating the question-answer pair, which when combined with the full articles, yields a single \synqa training data sample.

\subsection{Dialogue context attributon}\label{app:prompts-dialogue}

\begin{prompt}
\textbf{SYSTEM PROMPT}

You are an AI assistant that generates structured question-answer pairs based on a passage. Your goal is to create meaningful, factual, and reasoning-based questions that require connecting multiple sentences.

Follow these strict guidelines:

- Format the output as a \textbf{valid JSON array}, where each item has:

  - \texttt{"question"}: A clear, concise question.
  
  - \texttt{"answer"}: A short, factual response.
  
  - \texttt{"sentence\_numbers"}: A list of integers pointing to \textbf{all} relevant supporting sentences.

- \textbf{Ensure questions are generated in a random sentence order} (not sequential).

- Some questions \textbf{must reference multiple sentences} for reasoning.

- Some sentences should be \textbf{reused} across multiple questions.

- \textbf{Later questions should rely on earlier information} and use pronouns or indirect references to maintain logical flow.

- Introduce a mix of \textbf{fact-based, causal, and inference questions}.

- Avoid introducing \textbf{information not present in the passage}.

- Ensure \textbf{all relevant sentences are cited} for each answer.

Your response must be \textbf{valid JSON} containing 5 to 10 question-answer pairs.
\end{prompt}

and the user prompt:

\begin{prompt}
\textbf{USER PROMPT}

Here is a passage:

Title: [Title of the passage]

0. [First sentence of the passage]

1. [Second sentence of the passage]

2. [Third sentence of the passage]

3. [Fourth sentence of the passage]

...

Generate structured question-answer pairs following these constraints:

- \textbf{Return output in JSON format only}: \texttt{[{"question": "...", "answer": "...", "sentence\_numbers": [..]}, ...]}

- Use \textbf{random sentence order}, not sequential.

- Some questions should require \textbf{multiple sentences}.

- Some sentences should be \textbf{reused} across different Q\&A pairs.

- \textbf{Later questions must reference earlier ones} using pronouns or indirect mentions.

- \textbf{Include a mix of question types}:

  - Factual questions that can be answered directly from the passage.
  
  - Causal questions that require understanding relationships between sentences.
  
  - Inference-based questions that require implicit reasoning.

- Ensure \textbf{sentence numbers fully cover the reasoning required}.

Return \textbf{only} JSON, with no extra text.
\end{prompt}
\section{Prompts for zero-shot models}\label{app:zero-shot-prompts}
In order to obtain context attributions with the instruction-tuned LLM \cite{Jiang2023Mistral7, Dubey2024TheL3}, we use the following prompts:

\begin{prompt}
\textbf{SYSTEM PROMPT}

You are an AI assistant that identifies the sentence(s) in a provided context document most relevant for answering a specific question. Your task is to select only the sentence(s) containing the explicit information needed to answer the question accurately, without adding extra context.
\end{prompt}

\begin{prompt}
\textbf{USER PROMPT}

Context Document:

[numbered sentences from the context]

Question: [query text]

Answer: [answer text]

Based on the context document, identify the sentence number(s) from the following choices: [list of numbers]. Select only the sentence(s) that contain explicit information needed to answer the question directly.

Answer only with the corresponding number(s) in parentheses, without additional explanation.
\end{prompt}
\section{User Study}
\label{app:user_study}

\begin{figure*}[t]
    \centering
    \includegraphics[width=0.8\linewidth]{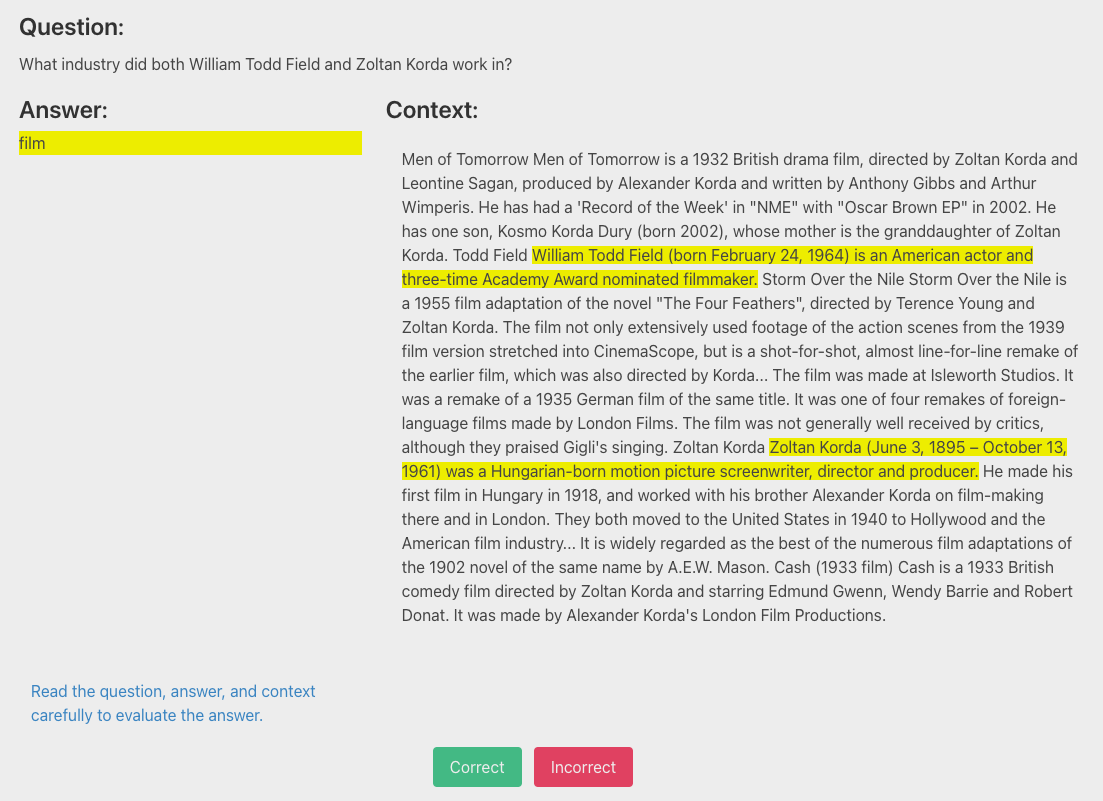}
    \caption{An example of the attribution scenario evaluated in our user study. Both the answer and the context attributions are highlighted to help the user verify the correctness of the answer. In the absence of highlights, the user is instructed to read the entire context. This example showcases a practical application of context attribution in real-world interactions with LLM-generated content.}
    \label{fig:user_interface}
\end{figure*}

An example of the attribution scenario evaluated in our user study. See Figure~\ref{fig:user_interface} for details.

\end{document}